\documentclass[smallextended]{svjour3}     

\usepackage{graphics}
\usepackage[round,colon]{natbib}
\usepackage{amsmath}
\usepackage{txfonts}
\usepackage[dvipdfm,colorlinks=true,citecolor=blue,linkcolor=blue]{hyperref}
\journalname{myjournal}
\begin{document}
\title{Macroscopic and microscopic statistical properties observed in blog entries}
\author{Yukie Sano \and Misako Takayasu}


\institute{
Department of Computational Intelligence and Systems Science,
Interdisciplinary Graduate School of Science and Engineering,
Tokyo Institute of Technology, 
4259 Nagatsuta-cho, Midori-ku, Yokohama 226-8502, Japan
\\
e-mail: sano@smp.dis.titech.ac.jp   
}

\date{Received: date / Revised version: date}
\maketitle
\begin{abstract}
We observe the statistical properties of blogs 
that are expected to reflect social human interaction. 
Firstly, we introduce a basic normalization preprocess 
that enables us to evaluate the genuine word frequency in blogs 
that are independent of external factors such as spam blogs, 
server-breakdowns, increase in the population of bloggers, and 
periodic weekly behaviors. 
After this process, we can confirm that small frequency words 
clearly follow an independent Poisson process as theoretically expected. 
Secondly, we focus on each blogger's basic behaviors. 
It is found that there are two kinds of behaviors of bloggers. 
Further, Zipf's law on word frequency is confirmed to 
be universally independent of individual activity types.
\end{abstract}
\section*{Keywords}
Blog analysis $\cdot$ Time series analysis $\cdot$ Zipf's law $\cdot$ Human dynamics

\section{Introduction}
\label{intro}
Blogs are a new kind of social communication medium 
in which personal opinions can be easily uploaded on the Web. 
A typical blog site is maintained by an individual or a small group. 
Blog users are called bloggers and they post blog ``entries'' 
that are freely written texts like those in diaries. 
These texts include opinions on movies, evaluations of purchased items and 
announcements of social events. 
Thus, each word in blog entries may reflect social phenomena. 
Search engine technologies have been developed 
to observe the details of blog entries automatically at high speeds. 
In this paper, we focus on the statistics of blogs from both macroscopic and microscopic perspectives.

\section{Data description}
\label{sec:1}
Using a search engine similar to ``Google Blog Search,''\footnote{http://blogsearch.google.com/}  
we analyzed Japanese blog databases that were collected from January 1st 2007 to 
December 31st 2008 by ``Dentsu Buzz Research''\footnote{http://www.dbuzz.jp/} 
For given keywords, observation period, and search area, 
the search engine automatically lists all entries that fulfill the condition. 
The search engine covers 20 major blog providers in Japan 
that host more than 10 million blog sites. 
The total number of observed entries is more than 610 million, 
and there are about 800,000 new entries uploaded daily on average. 
While we only focus on Japanese blogs in this paper, 
the share of Japanese blog sites is known to be largest, about 37\%, 
followed by English and Chinese blog sites for the year 2007 according to 
the report of Technorati,\footnote{http://technorati.com/}  
an internet search engine company for blogs (\citealt{technorati}).
In this paper, we firstly focus on the temporal change of word frequency on blog entries. 
Specially, we count number of blog entries including a target keyword at least once. 
Namely, if one blog entry includes the target keyword more than two times, 
we regard the number of blog entry as one.
We randomly choose words from a dictionary of Japanese morphological 
analysis,\footnote{http://chasen.naist.jp/} 
which is widely used in the field of natural language processing.

\section{Noise reductions}
\label{sec:2}
In this section, we introduce a basic procedure 
to evaluate the genuine word frequency in blogs independent of external factors. 

\subsection{Effect of spam blogs}
\label{sec:2-1}
While reading the collected blog entries, 
we easily find that there are blog entries 
that are obviously not generated by humans. 
For example, there are cases of blog 
entries' texts comprising of 
a meaningless sequence of words, copied articles 
from major internet news articles or simply repeated 
advertisement keywords. 
Further, some entries contain sexual or violent content 
that lead to a paid-membership site. 
Collectively, these examples are called \textit{spam} blogs. 
Some spam blogs are created with the intention 
to enhance their ranking at sites such as PageRanks (\citealt{page}). 
A large amount of spams is generated daily 
and it causes heavy fluctuations in word frequencies.

In the study of blog analysis, spam 
is attracting considerable interest, and thus, various methods for the  
detection of spams have been developed (\citealt{somu,narisawa,sato}). 
In the search engine of Dentsu Buzz Research, 
the following spam filters are installed:

\begin{itemize}
\item \textit{word salad}: Blog contents are a mixture of 
seemingly meaningful words that together signify nothing.
\item \textit{copy \& paste}: Blog contents are automatically or manually excerpted from other sources.
\item \textit{template}: Blog entry comprises template sentences and fixed keywords.
\item \textit{multi post}: Identical blog entries are posted to different blog sites.
\item \textit{adult and gamble}: Blog entry contains adult or gambling contents.
\end{itemize} 

In this paper, we use these spam filters to categorize spam and normal blogs. 
As a result of benchmark testing of the filter for 200 blog entries, 
the total detection accuracy was 83\%, 
about 40\% of the collected blog entries being categorized into spam.

\begin{figure}
\center
\scalebox{0.6}{\includegraphics{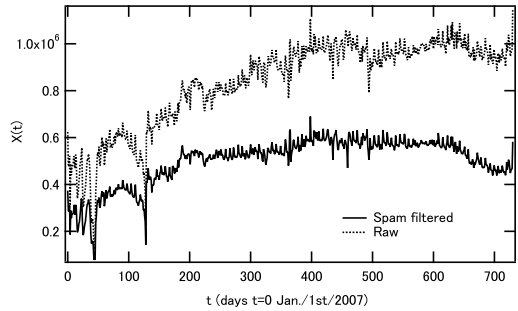}}
\caption{Time series of whole collected blog entries with spam filtered $X(t)$ and 
time series without filtered (dashed line)}
\label{fig:spam}       
\end{figure}

\subsection{Effect of system maintenance and population growth}
\label{sec:2-2}
Since blogs are supported by computer systems, 
there is a possibility that some blog servers suddenly 
stop working because of maintenance or hardware replacement, and thus, 
there may be a sudden decrease of word 
frequency for a period.
Moreover, there is a tendency that the total number of 
blog sites increase almost monotonically (\citealt{somu}). 
Hence, the average number of appearances of any word tends to increase in a non-stationary way.
Here, we introduce a procedure 
for adjusting these external or systemic non-stationary effects.

For a given time series of flux fluctuation, 
Menezes and Barab\'asi introduced a method 
of separating external noise effect from internal contributions 
in an open system of complex network model (\citealt{menezes}). 
Utilizing the fact  
that flux time series comprise  
independent small parts of fluxes, 
they computed the share of the small parts of fluxes 
in the entire range of collected fluxes. 
With regard to their mathematical models, 
the method works successfully by separating external noises from 
the time series. However, 
in this case, we cannot assume that each blogger acts independently. 
Therefore, we introduce a new revised method for 
the separation of internal and external fluctuations.

\begin{figure}
\center
\scalebox{0.7}{\includegraphics {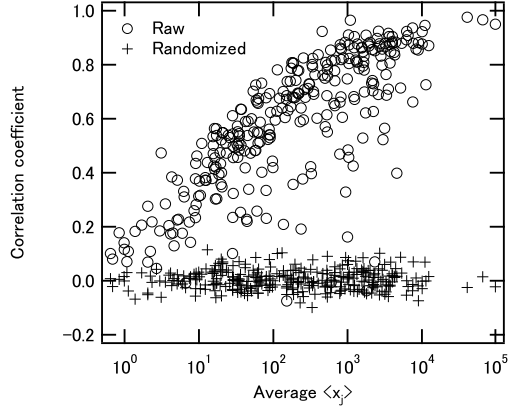}}
\caption{
Comparison of correlation between the average value of frequency $\langle x_j \rangle$ 
and correlation $C_j$ (circle) and $C'_j$ (cross). }
\label{fig:correl} 
\end{figure}

\begin{figure}
\center
\scalebox{0.6}{\includegraphics{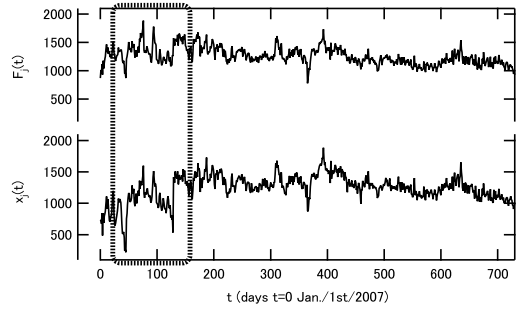}}
\caption{An example of a word with high frequency (keyword:``if'') 
before and after the normalization ($\langle F_j\rangle=1241.3$)} 
\label{fig:norm} 
\end{figure}

For words with low frequency, we assume that bloggers pay 
little attention to these words 
and the blog entry numbers may not be significantly affected 
by external factors. 
For words with high frequency, 
we conjecture that bloggers focus on 
appearance numbers, and thus, these words are significantly 
affected by external factors. 
Based on these assumptions, we discern 
that the contribution of external factors 
depends on the average value of word frequency.
For any given keyword $j$, 
we calculate the average value of daily blog entries time 
series $x_j(t)$ and the correlation coefficient between $x_j(t)$ and 
the time series of whole collected blogs through spam filters, $X(t)$, 
as shown in Fig.~\ref{fig:spam},
\begin{equation}
C_j=\frac{\sum_{t=0}^T\left(( X(t)- \langle X \rangle)\right)\left(x_j(t)-\langle x_j \rangle \right)}
{\sqrt{\sum_{t=0}^T\left( X(t)-\langle X \rangle \right)^2}\sqrt{\sum_{t=0}^T\left( x_j(t)-\langle x_j \rangle \right)^2}},
\label{eq:1}
\end{equation}
where $\langle X \rangle \equiv \frac{1}{T+1}\sum_{t=0}^TX(t)$ and 
$\langle x_j \rangle \equiv \frac{1}{T+1}\sum_{t=0}^Tx_j(t)$.
As shown in Fig.~\ref{fig:correl}, we confirm 
that a clear positive correlation exists between $\langle x_j \rangle$ and 
$C_j$ compared with the case that the blog timestamps
are randomly shuffled $C'_j$. 
Once we obtain $C_j$ from $\langle x_j \rangle$, 
we define a new normalization of the time series as follows.
\begin{equation}
F_j(t)=C_j\left( \frac{x_j(t)}{X(t)} \langle x \rangle \right)+(1-C_j)x_j(t)
\label{eq:2}
\end{equation}
As shown in Fig.~\ref{fig:correl}, for a small 
value of $\langle x_j \rangle$, the value of $C_j$ is also small. 
In this case, the first term of Eq.(\ref{eq:2}) can be neglected and we have 
$F_j(t)\thickapprox x_j(t)$. 
On the other hand, for a large value of $\langle x_j \rangle$, 
the second term in the right hand side of Eq.(\ref{eq:2}) 
becomes negligible and we have $F_j(t) \thickapprox \frac{x_j(t)}{X(t)}\langle X \rangle$.
An example demonstrating the effect of this normalization is shown in Fig.~\ref{fig:norm}. 
We confirm that systematic fluctuations are reduced from the time series (Fig.~\ref{fig:norm})

\subsection{Effect of weekly period}
\label{sec:2-3}
There are words that exhibit clear periodic behaviors 
depending on the day of week. 
For example, ``hospital'', ``office'', or ``school'' are typical words 
that appear more frequently on weekdays. 
For the purpose of flattening such a weekly period, 
we sum up the number of appearances of words for each day of the week, $N(k)$, 
$k=0,1,...6$, where $k=0$ means Sunday, $k=1$ means Monday, etc.
Then, the time series of word frequency, $F_j(t)$ is normalized by the following way:
\begin{equation}
\overline{F}_j(t) \equiv \frac{F_j(t)}{N_j \left(t \bmod 7 \right)}\frac{N_j}{7},
\label{eq:3}
\end{equation}
where $N_j \equiv \sum_{k=0}^6 N_j(k)$ is the total number of blog entries.
Note that for most words, such week dependence normalization is not necessary.

\begin{figure}
\center
\scalebox{0.6}{\includegraphics{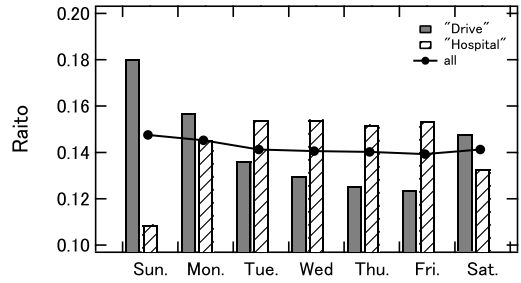}}
\caption{Ratio for the day of the week; keywords are ``hospital'' and ``drive''}
\label{fig:week} 
\end{figure}
\subsection{Result of noise reduction}
\label{sec:2-4}
\begin{figure}
\scalebox{0.5}{\includegraphics{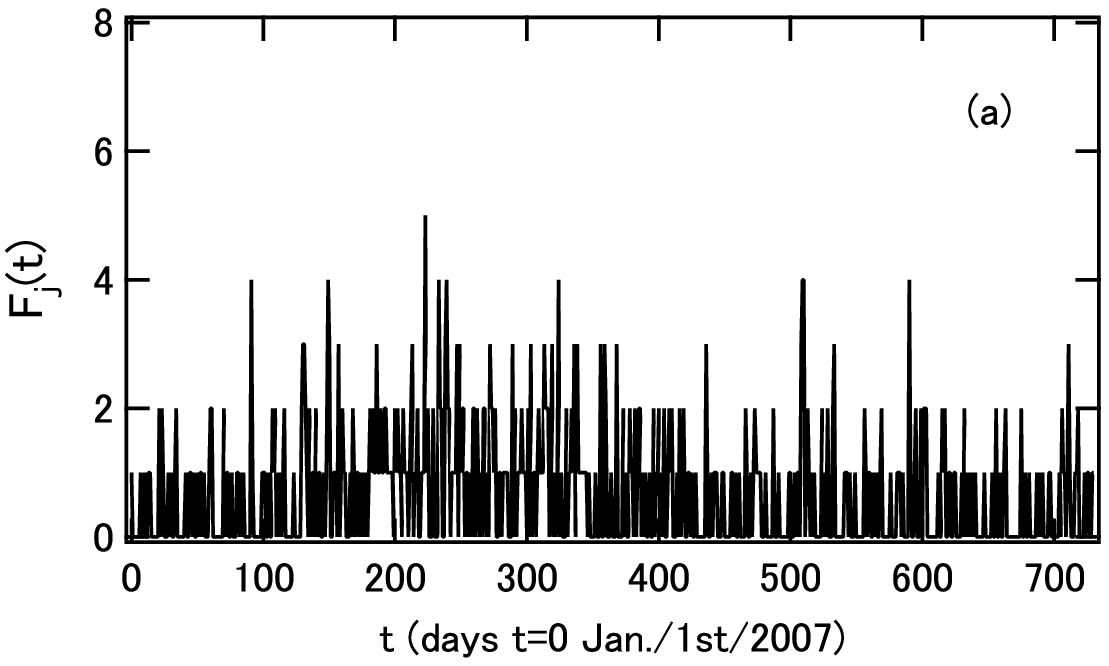}}
\scalebox{0.5}{\includegraphics{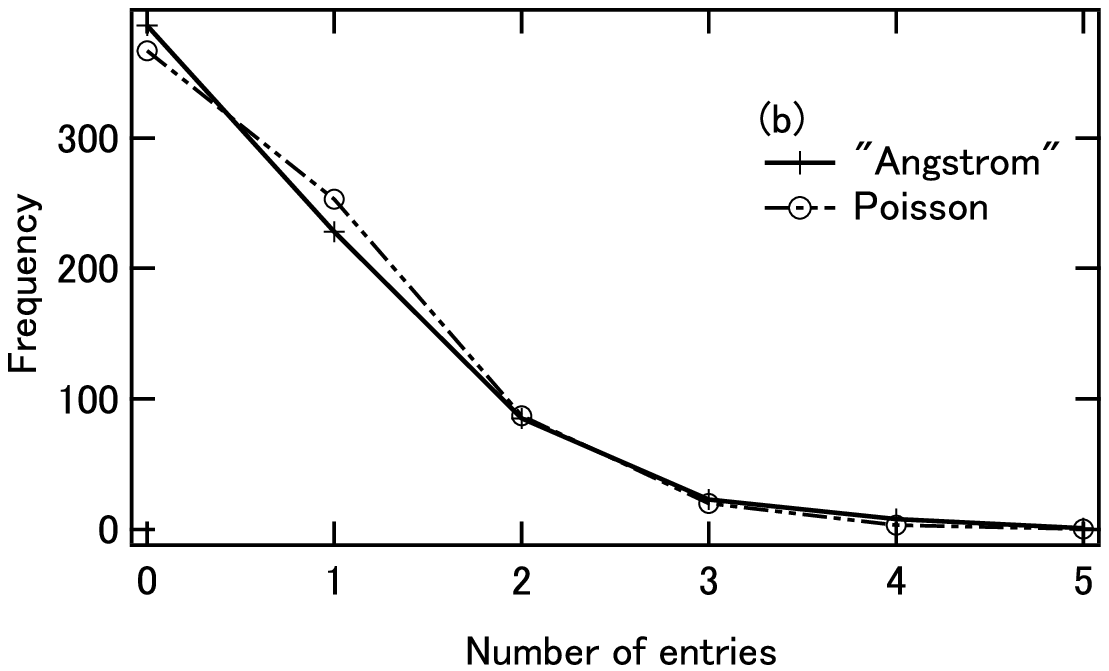}}
\caption{An example of time series {\bf(a)} and frequency distribution {\bf(b)} 
for less frequency word ($\langle F_j\rangle=0.7$); keyword is ``angstrom''}
\label{fig:angstrom} 
\end{figure}

With the noise reduction procedures introduced in 
Sec. \ref{sec:2-1}-Sec. \ref{sec:2-3}, 
we confirm that systematic noises are removed and 
that the time series appears more stationary. 
For words with low frequency, specifically, words that appear less than 1 times per day on average, 
we confirm that autocorrelation is $0$ in 95\% significance level. 
and the distribution of intervals of appearance is checked 
to pass the statistical tests of Poisson distribution 
as demonstrated in Fig.~\ref{fig:angstrom}.
The result of $\chi$ square test shows that it is not rejected by 2.5\% significance level.
As a result of randomly selected 300 words from Japanese morphological 
analysis dictionary, only one word, ``Angstrorm'', passed the $\chi$ square test while remaining words appeared more than 
1 times per day on average.

In a pioneering study of the basic statistics of blogs, while \citet{ausloos} 
reported that the Poissonian hypothesis is always rejected, 
they did not apply systematic noise reduction procedures. 
For words with high frequency, we generally find a clear deviation 
from the simple Poisson process as already presented (\citealt{ausloos}), 
even these noise reduction processes are fully applied. 
This implies that there is potential interaction among bloggers.

\section{Bloggers' individual properties}
\label{sec:3}
So far, we have discussed about word frequency in blogs 
from a macroscopic point of view. 
In this section, we focus on the individual properties of bloggers,  
which are expected to form the base for the development of agent-based modeling of blogs.

\subsection{Intervals of posting}
\label{sec:3-1}
\begin{figure}
\center
\scalebox{0.8}{\includegraphics{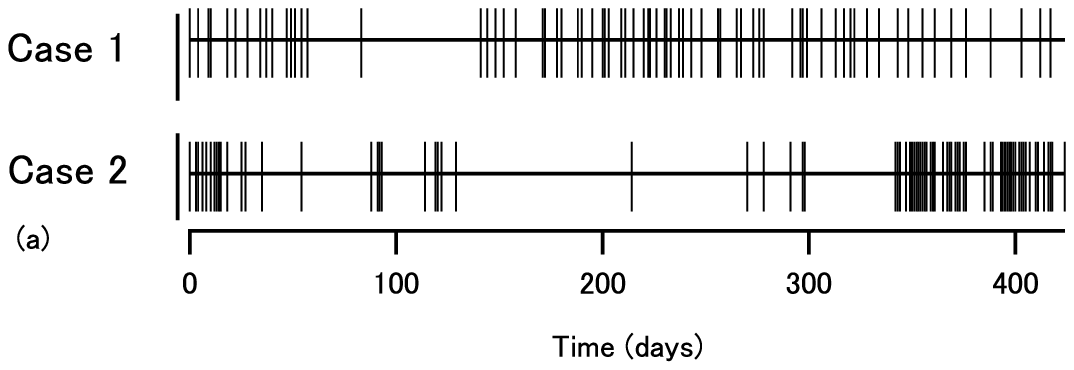}}\\
\scalebox{.8}{\includegraphics{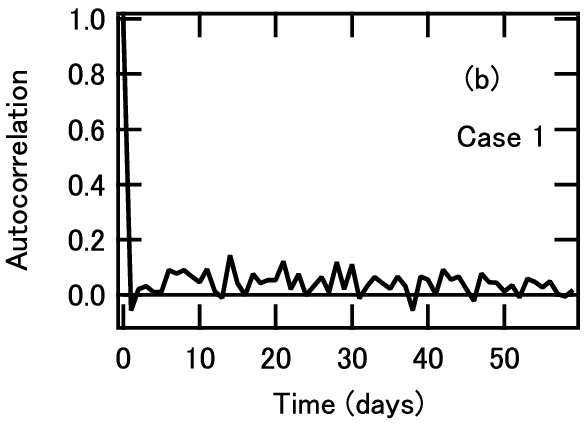}}
\scalebox{.8}{\includegraphics{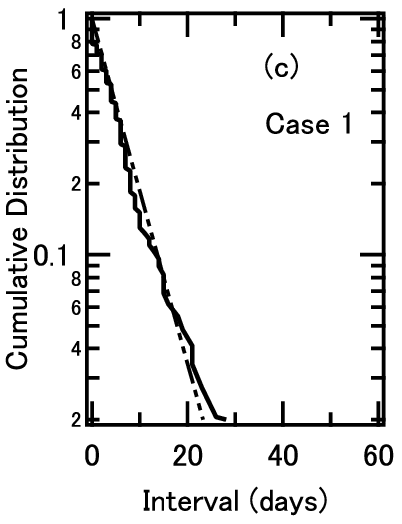}}\\
\scalebox{.8}{\includegraphics{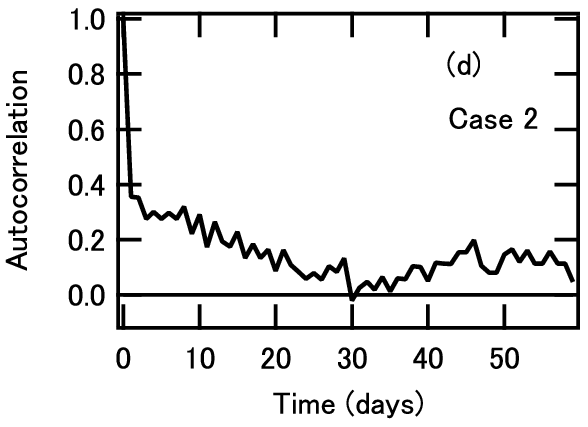}}
\scalebox{.8}{\includegraphics{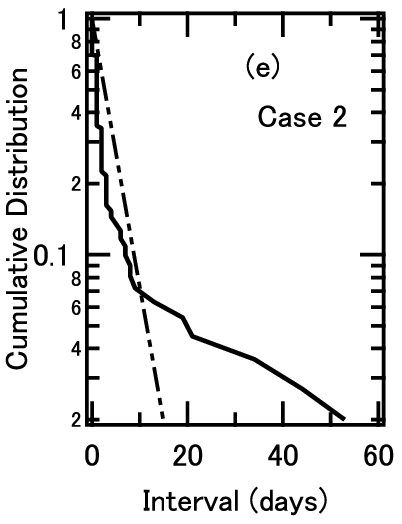}}
\caption{Comparison of two cases of posting blog entries {\bf (a)}; 
Case 1 is well approximated by Poisson process and 
Case 2 reveals non-trivial correlation;
Autocorrelation function {\bf (b), (d)}; and 
cumulative distribution of posting interval {\bf (c), (e)}}
\label{fig:id} 
\end{figure}

In this section, we focus on the bloggers' behaviors 
of posting blogs. 
We analyze bloggers' data in which individual bloggers' 
entries are recorded with the time stamp of precision in second   
from November 1st 2006 to March 31st 2009. 
If a blogger's behavior is approximated by an independent Poisson process, 
the distribution of intervals is approximated 
by an exponential function and the autocorrelation is almost $0$. 
From this theoretical viewpoint, we categorize the bloggers into two cases: 

\begin{itemize}
\item \textit{Case 1}: Poissonian bloggers
\item \textit{Case 2}: Non-Poissonian bloggers
\end{itemize}
In Fig.~\ref{fig:id}, we show two typical examples 
belonging to these two cases. 
In Case 1, the observed time series of postings 
(the top figure of Fig.~\ref{fig:id}a) is characterized 
by a quick decay of the autocorrelation function (Fig.~\ref{fig:id}b) 
and the distribution of intervals is well approximated 
by an exponential function (Fig.~\ref{fig:id}c). 
Therefore, an independent Poisson process can form the 
base of the behavior for such bloggers. 
On the other hand, in Case 2, 
the occurrence time series clearly shows clustering 
(the bottom figure of Fig.~\ref{fig:id}a), 
and the autocorrelation decays slowly (Fig.~\ref{fig:id}d). 
Furthermore, the interval distribution has a fat-tail 
that is approximated by a power law (Fig.~\ref{fig:id}e). 
From this example, we find that a non-Poissonian blogger 
possesses strong memory in that once he (or she) posts an entry, 
he (or she) tends to continue posting entries. 
In this analysis, we classify 110 bloggers into these two cases. 
There are 10 Poissonian bloggers and 
the remaining 100 bloggers are categorized into the non-Poissonian case 
by Kolmogorov-Smirnov test applied to the sequence of posting time intervals. 
We also confirm that autocorrelation functions of 
Poissonian bloggers are always within 95\% confidence bands. 
Additionally, we also checked 9753 bloggers during the observation period 
from November 1st 2006 to July 10th 2009 by the Kolmogorov-Smirnov test, 
and we confirmed that 1089 (about 11\%) are categorized into the Poissonian bloggers.
In contrast to the keyword appearance described 
in Sec. \ref{sec:2-4}, the behaviors of bloggers 
can not be simply categorized by the average number of entries, 
that is, there are both kinds of bloggers in any posting rate groups.

\subsection{Individual word frequency}
\label{sec:3-2}
In this section, we investigate the frequency of words in individual blog entries. 
The study of the frequency of words started in the 1930s 
by the linguist Zipf who counted the number of appearance of words 
in various documents. 
He determined the rank by sorting the words with respect to the frequency and 
found an empirical law that the frequency of 
a word is approximately proportional to the inverse of its rank (\citealt{zipf}). 
This old law, generally called ``Zipf's law,'' 
still attracts considerable interest among scientists of various fields 
because it is applicable not only to linguistic problems 
but also to a wide variety of phenomena 
such as the incomes of companies (\citealt{takayasu}) 
and the abundances of expressed genes distributions (\citealt{kaneko}). 

\begin{figure}
\center
\scalebox{0.6}{\includegraphics{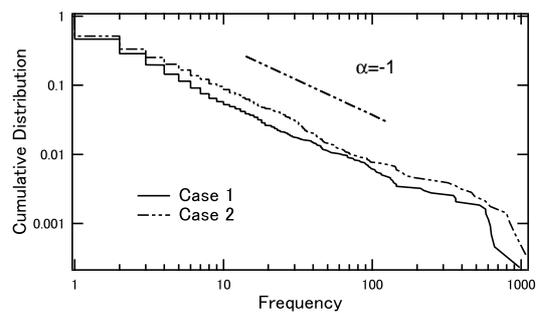}}
\caption{Word frequency distribution of different bloggers. 
The guideline follows a power law with an exponent of $-1$}
\label{fig:zipf} 
\end{figure}

In Fig.~\ref{fig:zipf}, word frequency distributions are plotted 
for both Poissonian and non-Poissonian bloggers, 
and we can find that Zipf's law holds in both cases. 
High frequency words are mainly postpositional particles 
and auxiliary verbs that commonly appear in all blog sites. 
Further, some topical keywords also appear frequently in each blogger's entries. 
In the case of words with low frequency, there are no common words 
and the words depend on each blogger's characteristics. 
It is noteworthy that even under such non-uniformity, Zipf's empirical law 
holds for each blogger's entries.

\section{An application of blogs}
\label{sec:4}
\begin{figure}
\begin{center}
\scalebox{0.5}{\includegraphics{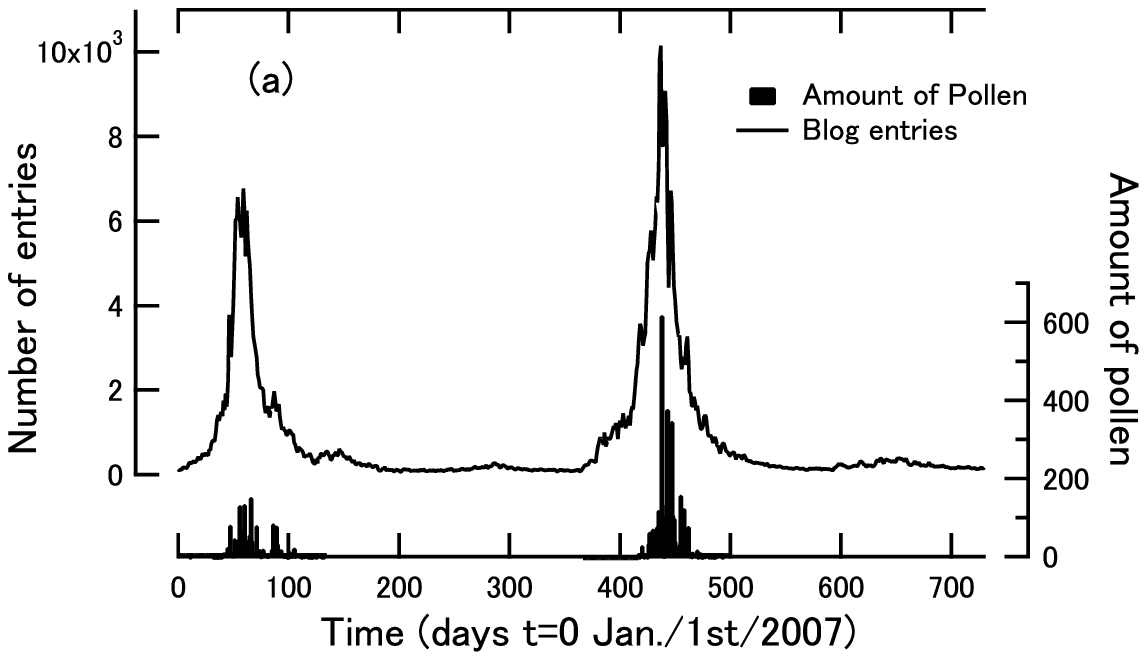}}
\scalebox{0.4}{\includegraphics{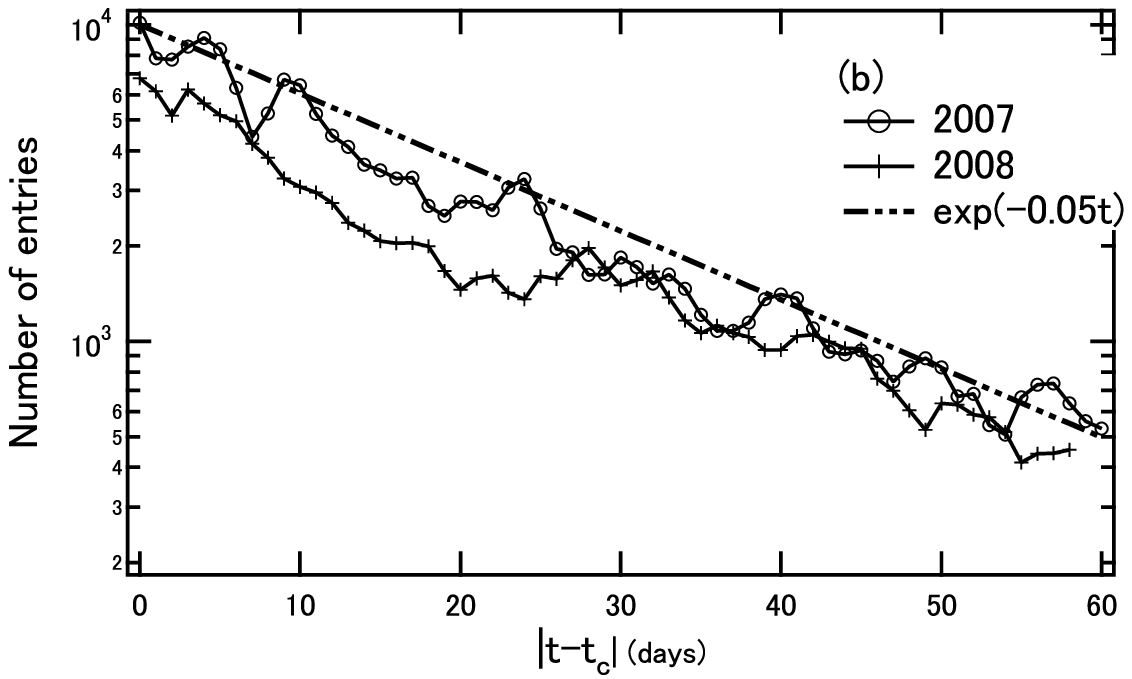}}
\caption{Time series including word ``pollen''  
and the amount of airborne pollen {\bf (a)}. 
Time series of after peak $t_c$ in semi-log scale {\bf (b)}}
\label{fig:pollen} 
\end{center}
\end{figure}

Finally, we introduce an example 
showing that blogs can efficiently capture certain kinds of 
social phenomena. 
Special keywords such as ``flu'' and ``pollen allergy'' 
appear periodically every year with sharp peaks as shown in Fig.~\ref{fig:pollen}a. 
We find that the number of blog entries including the keyword ``pollen'' 
is closely associated with the amount of airborne pollen in Tokyo.\footnote{http://kafun.jaanet.org/} 
Furthermore, this sharp rise and decay is well approximated 
by exponential functions as plotted in Fig.~\ref{fig:pollen}b. 
A recent paper by \citet{ginsberg} 
introduced a method of detecting influenza epidemics 
by using large numbers of Google search queries to track influenza-like illness. 
There is a possibility that blogs can also be used 
as an observation tool for the development of epidemics or allergy. 

\section{Discussion and conclusion}
\label{sec:5}
In this paper, we proposed a basic preprocess 
for the separation of external systematic noises from the time series of blog entries. 
After this normalization procedure, we confirmed that 
as theoretically expected, the appearance of 
low frequency words clearly follows a Poisson process. 
With regard to words with high frequency, 
the Poissonian assumption does not hold in any case, 
implying that existence of a strong non-trivial correlation among those words. 
We focused on each blogger's behavior of posting blog entries and 
found that bloggers can be categorized into two cases: 
Poissonian and non-Poissonian bloggers. 
About 20\% of bloggers belong to the Poissonian case 
in which basic behaviors can be modeled by an independent Poisson process. 
The rest of the bloggers tend to behave in an intermittent manner 
with strong memory effect. 
In any case, the word frequency follows Zipf's law for each individual.
 
We expect that these basic results will play an important role 
in the construction of a microscopic agent-based model of bloggers 
in the near future. One of the targets of such a microscopic model 
will be the explanation of the macroscopic behaviors related to the exponential 
rise and decay of commonly used keywords such as ``pollen'' shown in Sec. \ref{sec:4}.

\section*{Acknowledgement}
The authors are grateful to the corporations of Dentsu Inc. and Hottolink Inc. 
for providing blog data.
%

\end{document}